\documentclass[conference]{IEEEtran}
\IEEEoverridecommandlockouts
\usepackage{amsmath,amssymb,amsfonts}
\usepackage{algorithmic}
\usepackage{graphicx}
\usepackage{textcomp}
\usepackage{xcolor}
\usepackage[lined,ruled,vlined]{algorithm2e}
\usepackage[nolist,nohyperlinks]{acronym}
\def\BibTeX{{\rm B\kern-.05em{\sc i\kern-.025em b}\kern-.08em
    T\kern-.1667em\lower.7ex\hbox{E}\kern-.125emX}}
\begin{document}

\title{A Deep Multi-Modal Cyber-Attack Detection in Industrial Control Systems
}

\author{\IEEEauthorblockN{Sepideh Bahadoripour}
\IEEEauthorblockA{\textit{Schulich School of Engineering} \\
\textit{University of Calgary}\\
Calgary, Canada \\
sepideh.bahadoripour@ucalgary.ca}
\and
\IEEEauthorblockN{Ethan MacDonald}
\IEEEauthorblockA{\textit{Schulich School of Engineering} \\
\textit{University of Calgary}\\
Calgary, Canada \\
ethan.macdonald@ucalgary.ca}
\and
\IEEEauthorblockN{Hadis Karimipour}
\IEEEauthorblockA{\textit{Schulich School of Engineering} \\
\textit{University of Calgary}\\
Calgary, Canada \\
hadis.karimipour@ucalgary.ca}
}

\maketitle

\begin{abstract}
The growing number of cyber-attacks against Industrial Control Systems (ICS) in recent years has elevated security concerns due to the potential catastrophic impact. Considering the complex nature of ICS, detecting a cyber-attack in them is extremely challenging and requires advanced methods that can harness multiple data modalities. This research utilizes network and sensor modality data from ICS processed with a deep multi-modal cyber-attack detection model for ICS. Results using the Secure Water Treatment (SWaT) system show that the proposed model can outperform existing single modality models and recent works in the literature by achieving 0.99 precision, 0.98 recall, and 0.98 f-measure, which shows the effectiveness of using both modalities in a combined model for detecting cyber-attacks.
\end{abstract}

\begin{IEEEkeywords}
industrial control system, multi-modal, cyber-attack detection, machine learning, deep learning, time-series.
\end{IEEEkeywords}
\begin{acronym}
    \acro{AE}{AutoEncoder}
     \acro{AB}{Adaboost}
     \acro{AI}{Artificial Intelligence}
     \acro{ANN}{Artificial Neural Network}
     \acro{BCE}{Binary Cross Entropy}
     \acro{BP}{Back Propagation}
     \acro{CE}{Cross Entropy}
     \acro{CERT}{Computer Emergency Response Team}
     \acro{CMRI}{Complex Malicious Response Injection}
     \acro{CNN}{Convolutional Neural Network}
     \acro{CPS}{Cyber-Physical System}
     \acro{CVSS}{Common Vulnerability Scoring System}
     \acro{DARPA}{Defense Advanced Research Projects Agency}
     \acro{DBN}{Deep Belief Network}
     \acro{DCS}{Decentralized Control System}
     \acro{DNN}{Deep Neural Network}
     \acro{DoS}{Denial of Service}
     \acro{DRL}{Deep Reinforcement Learning}
     \acro{DT}{Decision Tree}
     \acro{ELM}{Extreme Learning Machine}
     \acro{FCM}{Fuzzy C-Means}
     \acro{FL}{Federated Learning}
     \acro{FN}{False Negative}
     \acro{FP}{False Positive}
     \acro{GAN}{Generative Adversarial Network}
     \acro{GB}{Gradient Boosting}
     \acro{GRU}{Gated Recurrent Unit}
     \acro{HMI}{Human-Machine Interface}
     \acro{HMM}{Hidden Markov Model}
     \acro{ICS}{Industrial Control System}
     \acro{IDS}{Intrusion Detection System}
     \acro{IED}{Intelligent Electronic Device}
     \acro{IF}{Isolation Forest}
     \acro{IIoT}{Industrial Internet of Things}
     \acro{IoT}{Internet of Things}
     \acro{IP}{Internet Protocol}
     \acro{IT}{Information Technology}
     \acro{KNN}{K-Nearest Neighbors}
     \acro{LAD}{Logical Analysis of Data}
     \acro{LR}{Logistic Regression}
     \acro{LSTM}{Long Short-Term Memory}
     \acro{MCC}{Matthews Correlation Coefficient}
     \acro{MDP}{Markov Decision Process}
     \acro{MFCI}{Malicious Function Command Injection}
     \acro{ML}{Machine Learning}
     \acro{MCSI}{Malicious State Command Injection}
     \acro{MPCI}{Malicious Parameter Command Injection}
     \acro{MRE}{Mean Reconstruction Error}
     \acro{MSE}{Mean Square Error}
     \acro{NADAM}{Nesterov-accelerated Adaptive Moment Estimation}
     \acro{NB}{Na\"ive Bayes}
     \acro{NMRI}{Na\"ive Malicious Response Injection}
     \acro{OCSVM}{One-Class Support Vector Machine}
     \acro{OT}{Operational Technology}
     \acro{PCA}{Principal Component Analysis}
     \acro{PLC}{Programmable Logic Controller}
     \acro{PMU}{Phasor Measurement Unit}
     \acro{PPO}{Proximal Policy Optimization}
     \acro{RBF}{Radial Basis Function}
     \acro{Recon}{Reconnaissance}
     \acro{ReLU}{Rectified Linear Unit}
     \acro{RF}{Random Forest}
     \acro{RL}{Reinforcement Learning}
     \acro{RNN}{Recurrent Neural Network}
     \acro{SAE}{Stacked Autoencoder}
     \acro{SCADA}{Supervisory Control And Data Acquisition}
     \acro{SQL}{Structured Query Language}
     \acro{SVM}{Support Vector Machine}
     \acro{SWaT}{Secure Water Treatment}
     \acro{TN}{True Negative}
     \acro{TP}{True Positive}
     \acro{US}{United States}
     \acro{WAN}{Wide Area Network}
    \end{acronym}

\section{Introduction}
\label{sec:intro}
\ac{ICS}s are crucial for monitoring and controlling critical infrastructure such as smart grids, oil and gas industry, and transportation. Traditionally, \ac{ICS}s were placed on isolated communications systems to keep them safe from cyber-attacks, where they were relatively inconspicuous and unknown to most attackers \cite{SAKHNINI2021101394}.

The connection between \ac{ICS} wide area networks facilitates the online access, monitoring, and control of the systems remotely. The drawback of wide area network integration is increased attack surfaces in ICS and increased vulnerability to cyber-criminals \cite{Zolanvari2019}. The Stuxnet campaign that targeted Iranian centrifuges for nuclear enrichment in 2010, and caused severe damage to the equipment is one high-profile example \cite{Zhang2019}. Several other prominent examples include: cyber-attacks affecting three U.S. gas pipeline companies in 2018 \cite{Zhang2019}, and two U.S. water treatment systems in 2021 \cite{Collier2021}. Due to the complexity and sensitivity of the {ICS} environment, a compromise in the system may lead to severe danger to human life or the environment. Moreover, the growing number of cyber-attacks against \ac{ICS} in recent years motivates the need for timely security solutions for these systems \cite{9645267}.

Although security solutions developed for \ac{IT} are mature, they are not sufficient due to differences between \ac{ICS} and \ac{IT} \cite{Yan2019}. While the \ac{IT} networks mainly focus on managing the high throughput of a network, \ac{ICS} networks have to function more towards carrying out tasks reliably and punctually. When failure occurs, the \ac{ICS} network cannot merely be rebooted, like a typical computer network \cite{Singh2020}. In addition, \ac{ICS} introduces more security vulnerability due to the tight integration between the controlled physical environment and the cyber system \cite{Paridari2018}. Compared to the IT protocols, ICS protocols, including Modbus or DNP3, suffer from common security issues such as the absence of authentication, lack of protection or security measures for data transferring over the link, and insufficient control measures to avoid default broadcast approaches \cite{Singh2020}. 

Considering the vital role of \ac{ICS} in critical infrastructural operations, various methods for cyber-threat detection and classification have been proposed in the literature to address the above-mentioned challenges. Model-based solutions, such as variants of state estimation techniques and statistical-based models, have been suggested in \cite{ShuguangCui2012}. Additionally, the use of Kalman filters for measurement estimation is also proposed to detect cyber-attacks \cite{Kurt2018}. However, such model-based techniques are difficult to adapt to larger stochastic systems and various threats. Intelligent systems are thus proposed for detecting and classifying cyber-attacks \cite{7762123}.

The current work proposes a multi-modal cyber-attack detection model to analyze sensor and network modalities of an \ac{ICS} and detect cyber-attacks based on their joint abstract representation. Due to the different structures of sensor and network data, a complex deep neural network includes a partially connected stacked \ac{LSTM} and neural network model, connected to a fully connected fusion network is proposed in this research. The proposed model is able to analyze each modality separately and then combine the key findings of each modality automatically using a deep neural network with a single training.

The rest of this paper is organized as follows. Section \ref{sec:LR} introduces some works in the literature. Section \ref{sec:pm} explains the proposed model, which is followed by the experimental setup in Section \ref{sec:exp}. Section \ref{sec:res} presents the evaluation results of the proposed model and compares it with some single modality models in the literature. Finally, Section \ref{sec:conc} concludes the paper.

\section{Literature Review}
\label{sec:LR}
\ac{ML}-based attack detection techniques work based on a moving target to continually evolve and learn new vulnerabilities rather than identifying the attack signatures or the regular pattern of the network \cite{W19}. There are different \ac{ML} algorithms available in the literature to detect cyber-attacks compromising data integrity \cite{karimipourIET}, availability \cite{Krawczyk2016}, and confidentiality \cite{Zolanvari2019}. \ac{ML}-based methods are categorized into two main categories, conventional and \ac{DNN}.
\subsection{Conventional Machine Learning-Based Cyber-Attack Detection}
In work by Zolanvari et al. \cite{Zolanvari2019}, the authors compared several traditional \ac{ML} algorithms, including \ac{KNN}, \ac{RF}, \ac{DT}, \ac{LR}, \ac{ANN}, \ac{NB}, and \ac{SVM} to detect cyber-attacks in a water storage system. Their evaluation showed that the \ac{RF} algorithm is the best model with a recall of 0.9744, the \ac{ANN} is the fifth algorithm with a recall of 0.8718, and the \ac{LR} is the worst algorithm with a recall of 0.4744. They also reported that the \ac{ANN} could not detect 12.82\% of the attacks but considered 0.03\% of the normal samples as an attack due to the imbalanced nature of the data. Based on the results, the authors considered many attack samples as normal samples without labeling so many normal samples as an attack happened for \ac{LR}, \ac{SVM}, and \ac{KNN}. Therefore, these \ac{ML} algorithms are sensitive to imbalanced data and unsuitable for \ac{ICS} attack detection.

In 2019, Sakhnini et al. \cite{sakhnini2019smart} proposed three feature selection methods, including binary cuckoo search, a genetic algorithm, and binary particle swarm optimization, which were analyzed to detect false data injection attack samples. The selected features were passed to \ac{SVM}, \ac{KNN}, and \ac{ANN} models for classification. The evaluation shows that the \ac{SVM} model outperforms the other classifiers in detecting cyber-attacks on three smart grid datasets.
Moreover, the evaluation illustrated that the genetic algorithm feature selection was the best at selecting a set of features among the feature selection methods.
In \cite{DAS2020101935}, the authors proposed a \ac{LAD} method to extract patterns/rules from the sensor data and used them to design a two-step anomaly detection system. They compared the proposed \ac{LAD} method with \ac{DNN}, \ac{SVM}, and \ac{CNN} models. The evaluations showed that the \ac{DNN} outperformed the \ac{LAD} method in the precision metric; however, the \ac{LAD} performed better in recall and f-measure. 
In \cite{al2020unsupervised}, a \ac{SAE}-based cyber-attack detection model for smart grid systems was proposed. This model was evaluated using three smart grid datasets and compared with conventional \ac{ML} approaches. The evaluations showed that the proposed method outperformed all other methods in all the metrics.
In 2022, Mittal et al. \cite{MITTAL202224} proposed a gravitational search algorithm-based clustering to analyze the \ac{ICS} data and detect cyber-attacks. Their model was evaluated on several datasets and compared with some other search-based algorithms in the literature.

\subsection{Deep Learning-Based Cyber-Attack Detection}
In \cite{7926429}, a deep cyber-attack detection method for real-time detection of false data injection attacks was proposed for smart grid systems. In this study, the effect of using different numbers of hidden layers on the model's performance is analyzed. The evaluations showed that the accuracy of a deeper model is higher than the shallower ones. 
Later work by Wang et al. \cite{wang2018} proposed a \ac{DNN} state estimation method for cyber-attack detection. In this method, the state estimation was done using an \ac{SAE}, and an attack was detected by comparing the estimated state with the actual one.
In \cite{Mariam}, the authors used autoencoders to detect cyber-attacks and restore injected data. Two types of autoencoders were used in this study. An undercomplete model was used for cyber-attack detection, and a denoising model was used to restore the attacked data. They compared this method to \ac{SVM}-based cyber-attack detection approach and showed that their proposed method outperformed them in attack detection, especially the previous unseen attack samples.
Moreover, in \cite{Karimipour2019}, the authors proposed a scalable deep cyber-attack detection for smart grids. This method was evaluated on a large-scale smart grid dataset and reported an accuracy of around 99\% for all scenarios.

Moreover, in 2021, Althobaiti et al. \cite{althobaiti2021intelligent} proposed a \ac{GRU}-based cognitive computing intrusion detection for \ac{ICS}. In their proposed method, first, the optimal subset of features was selected using a feature selection approach. Then, a \ac{GRU} model was applied to the selected features to detect cyber-attacks. Moreover, \ac{NADAM} optimizer was used to increase the model's accuracy.
In another manuscript, a novel deep representation learning approach was used to handle the challenge of imbalanced data in detecting cyber-attacks \cite{Jahromi2021}. This model was evaluated using two \ac{ICS} datasets and the results show the effectiveness of the proposed method.
In addition, in \cite{khan2022enhancing}, the authors proposed a \ac{LSTM} encoder-decoder model to detect cyber-attacks. In this method, a raw network traffic dataset was used, and a feature extraction method was used to extract features from the raw data. The extracted features were then passed through an encoder-decoder model and, the reconstruction error of this model is used to detect cyber-attacks.
This method was evaluated using an \ac{ICS} dataset that shows 97.95\% accuracy, 0.98 precision, 0.9663 recall, and 0.9789 f-measure on the gas pipeline dataset. 

In another study \cite{majidi2022fdi}, the authors proposed a \ac{DNN}-based method to detect false data injection attacks. The chi-squared test was first performed on the dataset to assign a weight to each feature based on its correlation with the output. The weighted features were then passed through a \ac{SAE} to generate new representations. Finally, the data in the latent space was used to build an ensemble of randomized \ac{DT}s. The majority voting method is used to detect cyber-attacks, and was evaluated on four power system datasets and compared with conventional \ac{ML} methods. The results illustrate that the DNN method outperforms the others in detecting cyber-attacks.
Recently, federated learning-based cyber-attack detection approaches have increased in popularity due to the ability to preserve privacy of \ac{ICS} data by training the models locally inside the clients and share the model parameters with a server to build a global model for all the clients \cite{jahromi2021deep,amir2}. In another study, Kravchik and Shabtai \cite{KravchikMoshe2021ECAD} proposed a lightweight deep model to detect \ac{ICS} cyber-attacks. They used 1D \ac{CNN} and undercomplete \ac{AE} to detect cyber-attacks using sensor data. Their experiments on the \ac{SWaT} dataset showed that their \ac{CNN} model achieved 0.90 precision, 0.87 recall, and 0.88 f-measure while their \ac{AE} model gained 0.96 precision, 0.93 recall, and 0.94 f-measure.

While the existing \ac{ML}-based approach can achieve acceptable results in detecting cyber-attacks in \ac{ICS}, to the best of the authors, none of them consider the multi-modality of \ac{ICS} environments in their solutions. To handle this gap, this work proposes a multi-modal cyber-attack detection component that used sensor and network modalities of \ac{ICS} environment to detect cyber-attacks more accurately.

\section{The Proposed Deep Multi-Modal Cyber-Attack Detection Model}
\label{sec:pm}
\begin{figure*}
	\centering
	\includegraphics[width=0.7\linewidth]{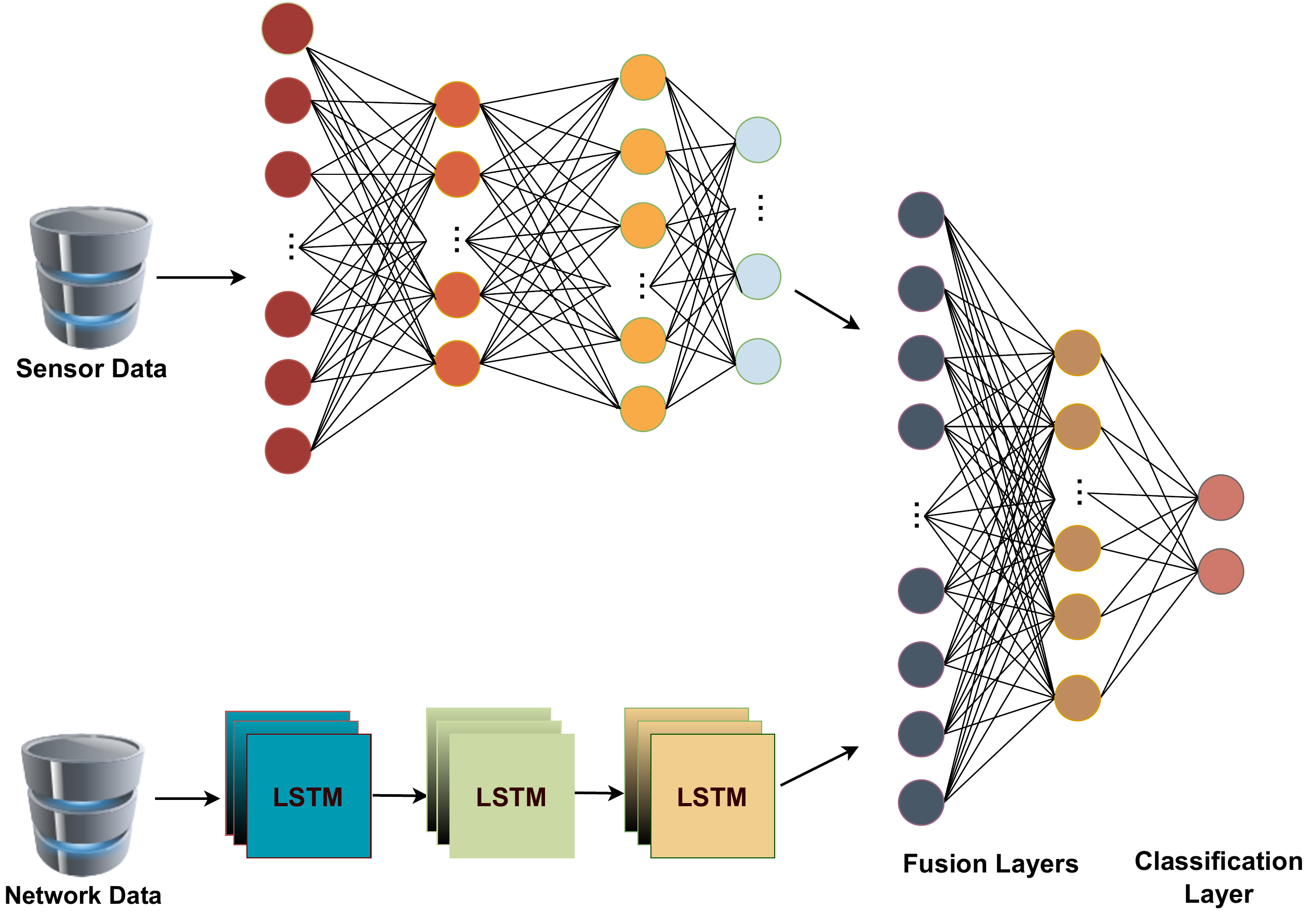}
	\caption{The proposed multi-modal cyber-attack detection model, including a fully connected deep model to analyze sensor modality, a stacked \ac{LSTM} model to analyze network modality, a fully connected deep model as a fusion model, and a single binary classification layer.}
	\label{fig:model}
\end{figure*}

In the proposed model, a multi-modal \ac{DNN}, including network and sensor modalities, was used to detect cyber-attacks in real time.
For each sample in sensor data, several network samples are available that show the transmitted packets in the \ac{ICS} network between \ac{IT} and \ac{OT} layers. A fully connected deep neural network is used to map the sensor data to a latent space, while a stacked \ac{LSTM} model is used to analyze the network data sequence and map it to the network latent space. These two new representations are then fused into a fully connected fusion network and finally classified using a single layer. Figure \ref{fig:model} shows the proposed multi-modal cyber-attack detection model. Assume $D = \{(x_S^{(i)}, x_N^{(i)}), y^{(i)}\}$ is the data where $x_S$ is a vector belongs to sensor modality, $x_N$ is a matrix of several network modality, $y$ is the label, and $i$ is the sample number which is $i\in (1,M)$ where $M$ is the number of samples in the dataset. The following subsections explain each part of the proposed model in detail.

\subsection{Sensor Model}
Sensor modality contains a vector of features for each sample in the dataset. The features of this modality (as will be discussed in section \ref{sec:exp}) are the sensor measurement and actuator status of the \ac{ICS}. A fully connected \ac{DNN} is proposed to analyze these features $x_S$ and map them to a sensor latent space $h_S$.
In the training step, the sensor part of each sample $x_S$ is passed to this model. Equation \ref{eq:3} shows the mapping of an input to a new representation using a single fully connected layer.
\begin{equation}
	h_S = \sigma (x_S, W)
	\label{eq:3}
\end{equation}
where $h_S$ is the new representation of sample $x_S$, $W$ is the weights of this layer, and $\sigma$ is the layer's activation function. The output of this layer is passed to the next layer to make a more abstract non-linear representation of the sensor data.

In this work, a four-layer fully connected \ac{DNN} is used for sensor modality. The output of the forth-layer $h_S$ is used as the final representation of the this modality. \ac{ReLU} is used as the activation function in this model ( \ref{eq:5}).
\begin{equation}
	ReLU(x) = max(0, x)
	\label{eq:5}
\end{equation}

\subsection{Network Model}
In each \ac{ICS}, several network packets are transmitted between different assets in the network at each time. So, for each sensor sample, several network samples are available. Therefore, a stacked \ac{LSTM} model is proposed to handle the transmitted packets $x_N$ and map them into the network latent space $h_N$.
In the training process, the network part of each sample $x_N$ is passed to this model. In an \ac{LSTM} model, the training is done based on the samples in a time-series. Equation \ref{eq:1} shows the mapping of the time-series sample $x_N$ to the hidden latent space $h_N$.
\begin{equation}
	h_N^t = f(Ux_N^t, Wh_N^{t-1})
	\label{eq:1}
\end{equation} 
where $h_N^T$ is the network latent space for the $t$-th time in the time-series, $U$ and $W$ are the \ac{LSTM} weights, and $x_N^t$ is the feature vectors of the $t$-th time-series of the network samples, and $f$ is the non-linear activation function of this model. Equation \ref{eq:2} shows the output of the proposed \ac{LSTM} model. 
\begin{equation}
	o_N^t = f(Vh_N^t)
	\label{eq:2}
\end{equation}
where $o_N^t$ is the output for the $t$-th element of the time-series, $V$ is the \ac{LSTM} weight, $h_N^T$ is the network latent space for the $t$-th time in the time-series, and $f$ is a non-linear activation function.

In the proposed network modality model, three \ac{LSTM} layers are stacked to form a stacked \ac{LSTM} model. In this situation, the outputs of each time-series element in an \ac{LSTM} (except the last one) are concatenated together to form a matrix and are used as a time-series input of the next \ac{LSTM} model. However, for the last \ac{LSTM}, only the final output is used as the network representation of the model $o_N$.

\subsection{Fusion and Classification Model}
After mapping the original features from each modality to a latent space, these two new representations are then passed to another fully connected deep model to fuse them. The output of this model is a joint representation of both modalities that consider the main abstract information in them. 

Sensor and network representations, $h_S$ and $o_N$, are then passed to this model to fuse the learned representations. Equation \ref{eq:4} shows this representation learning.
\begin{equation}
	h = \sigma((h_S, o_N), W)
	\label{eq:4}
\end{equation}
where $h$ is the joint representation, $h_S$ is the sensor representation generated from the sensor modality model, $o_N$ is the network representation generated from the network modality model, and $W$ is the weights of this model. This network is a two-layer fully connected neural network with \ac{ReLU} activation function (see equation \ref{eq:5}).

The joint representation is then passed through a single classification layer with binary cross entropy loss (see equation \ref{eq:6}) and softmax activation function (see equation \ref{eq:7}) to classify the samples as normal or attack, based on their sensor and network modalities, using equation \ref{eq:8}.
\begin{equation}
	\hat{y} = softmax(h\times W_c + b_c)
	\label{eq:8}
\end{equation}
where $\hat{y}$ is the predicted label, $h$ is the joint representation, $W_c$ is the classifier weights, and $b_c$ is the classifiers bias.
\begin{equation}
	L = \frac{-1}{M} \sum_{i=1}^{M} (y_ilog\hat{y_i} + (1-(y_i)log(1-\hat{y_i})))
	\label{eq:6}
\end{equation}
where $L$ is the binary cross entropy loss, $M$ is the number of samples, $y_i$ is the actual label of the $i$-th sample, and $\hat{y_i}$ is the predicted label for the $i$-th sample.
\begin{equation}
	softmax(s_i) = \frac{e^{s_i}}{\sum_{j=1}^{2} e^{s_j}}
	\label{eq:7}
\end{equation}

Algorithm \ref{alg:1} shows the algorithm of the proposed model.

\begin{algorithm}
	\SetAlgoLined
	\KwData{Multi-modal data $D = \{(x_S^{(i)}, x_N^{(i)}), y^{(i)}\}$ $i \in (1,M)$}
	\textit{Pre-processing}
	\begin{itemize}
		\item Impute missing values:\newline
		$x_{mod}^i [k] = \frac{1}{M} \sum_{j=1}^{M} x_{mod}^j [k], \newline if x_{mod}^j \neq None, \newline mod \in {Sensor, Network}$
		
		\item Data normalization:\newline
		$x_{mod}^{normalized} = \frac{x-min(x_{mod})}{max(x_{mod})-min(x_{mod})} \newline mod \in {Sensor, Network}$
		
	\end{itemize}

	\textit{Training:}\newline
	\For{Epoch e}{
		\begin{itemize}
			\item Pass $x_S$ through a fully connected \newline neural network:\newline
				$h_S = \sigma (x_S, W_S)$
			\item Pass $x_N$ through an stacked \ac{LSTM}:\newline
				$h_N^t = f(Ux_N^t, Wh_N^{t-1})$ \newline
				$o_N^t = f(Vh_N^t)$
			\item Pass the new representations through the fusion\newline model:\newline
				$h = \sigma((h_S, o_N), W)$
			\item Use a classification layer for cyber-attack \newline detection:\newline
				$\hat{y} = softmax(h\times W_c + b_c)$
			\item Use binary cross entropy to update the weights:\newline
				$L = \frac{-1}{M} \sum_{i=1}^{M} (y_ilog\hat{y_i} +\newline (1-y_i)log(1-\hat{y_i})))$
		\end{itemize}
		}
	\KwOut{Predicted labels for each sample $\hat{y}$}
	\caption{ The algorithm of the proposed multi-modal cyber-attack detection model.}
	\label{alg:1}
\end{algorithm}

\section{Experimental Setup}
\label{sec:exp}
To evaluate the proposed model, the \ac{SWaT} dataset \cite{Goh2017}, which includes both network and sensor modalities, was used. The \ac{SWaT} dataset is collected at the Singapore University of Technology from a water treatment system consisting of 449,920 samples. In this dataset, 87.9\% and 12.1\% were normal and attack samples, respectively. Sensor modality includes 51 features, each showing a sensor measurement or an actuator status. On the other hand, network
modality includes 16 features extracted from the transmitted packets in the \ac{ICS} network.

\subsection{Pre-Processing}
The mentioned data is pre-processed before passing it to the model. First, the mean value of each feature is used to impute the missing values. Then, equation \ref{eq:norm} is used to normalize the data between zero and one.
\begin{equation}
    x_{normalized} = \frac{x-min(x)}{max(x) - min (x)}
    \label{eq:norm}
\end{equation}

\subsection{Evaluation Metrics}
The proposed model is compared with several recent studies in the literature. Similar to other approaches, we used standard ML metrics to evaluate the performance of the proposed model. The basic \ac{ML} evaluation metrics are: 

	\begin{itemize}
		\item True Positive (TP)– Number of correctly classified samples as attack.  
		\item True Negative (TN)– Number of correctly classified samples as normal.  
		\item False Positive (FP)– Number of wrongly classified samples as attack.
		\item False Negative (FN)– Number of wrongly classified samples as normal.
	\end{itemize}
	
	Based on these basic metrics, more complicated metrics are defined that are used in this work, including:
	
	\begin{itemize}
	    \item \textbf{Precision} indicates the number of samples that are detected correctly as attack over total samples detected as attack (see equation \ref{eq:pre}).
	    
	    \item \textbf{Recall} indicates the number of samples that are detected correctly as attack over the total samples of the attack in the dataset (see equation \ref{eq:rec}).
	    
	    \item \textbf{F1-Score} is the harmonic value of precision and recall (see equation \ref{eq:f1}).
	\end{itemize}
	
	Due to the use of imbalanced data in this work, accuracy metric is not considered for comparison.
		\begin{equation}
		Precision = \dfrac{TP}{TP+FP}
		\label{eq:pre}
	\end{equation}
	
	\begin{equation}
		Recall = \dfrac{TP}{TP+FN}
		\label{eq:rec}
	\end{equation}
	
	\begin{equation}
		F1-Score = \dfrac{2 \times Precision \times Recall}{Precision + Recall}
		\label{eq:f1}
	\end{equation}

\section{Results and Discussion}
\label{sec:res}
The proposed multi-modal cyber-attack detection model was evaluated using an \ac{ICS} dataset consisting of a sensor and network modalities. The use of multiple modalities of data in the proposed model allows for a more comprehensive analysis of the ICS environment, which can improve the robustness and accuracy of cyber-attack detection. However, the utilization of different data modalities also poses certain challenges, such as the need to synchronize the datasets before training the model. This process can be time-consuming, especially when dealing with large datasets, and is considered the main limitation of the proposed model.

 Table \ref{tbl:res} shows the experimental results of evaluating the proposed model and its comparison with the single modality models and two existing works in the literature. As illustrated in this table, the proposed multi-modal model achieved a precision of 0.99, recall of 0.98, and f-measure of 0.98, which outperformed the other methods in all metrics. These results demonstrate the effectiveness of using both sensor and network modalities in a single model for cyber-attack detection in ICS environments.
  While the single modal models in Table \ref{tbl:res} gain a high precision, they cannot achieve a good recall due to the use of imbalanced data that caused a high number of evaded attacks from the model. However, the proposed model can handle this challenge and achieve a high recall that shows the effectiveness of using both modalities in a single model. On the other hand, LAD-ADS \cite{DAS2020101935} and CNN \cite{KravchikMoshe2021ECAD} models achieved a reasonable recall score by modifying the model that sacrificed the precision score, resulting in a higher number of false alarms of these models, which is handled in the proposed model.

\begin{table}[]
	\centering
	\caption{Results of the evaluation of the proposed multi-modal cyber-attack detection model and its comparison with single modality approaches.}
	\label{tbl:res}
	\renewcommand{\arraystretch}{1.50}
	\begin{tabular}{l|c|c|c}
		\multicolumn{1}{c|}{\textbf{Model}} & \textbf{Precision} & \textbf{Recall} & \textbf{f-measure} \\ \hline
		\textbf{Proposed Method} &\textbf{0.99} & \textbf{0.98} & \textbf{0.98} \\ 
		Sensor Model& 0.98 & 0.68 & 0.80 \\ 
		Network Model&0.95&0.63&0.75\\
		LAD-ADS \cite{DAS2020101935}&0.94&0.89&0.91\\
		CNN \cite{KravchikMoshe2021ECAD}& 0.0.90 & 0.87 & 0.88\\
            AE \cite{KravchikMoshe2021ECAD}& 0.96 & 0.93 & 0.94\\
	\end{tabular}
\end{table}

\begin{figure}
	\centering
	\includegraphics[width= .45\textwidth]{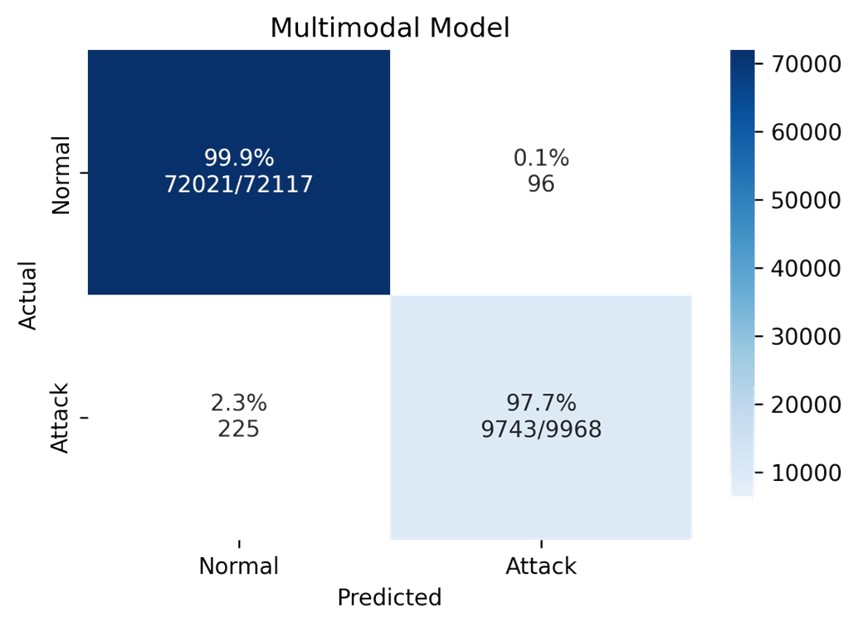}
	\caption{The confusion matrix of the proposed multi-modal cyber-attack detection model.}
	\label{fig:cm}
\end{figure}

Figure \ref{fig:cm} shows the confusion matrix of the proposed model. As illustrated in this figure, the proposed model only has 0.1\% of false alarms and 2.3\% of evaded attacks, which shows its good performance in detecting cyber-attacks. Having a low number of evaded attacks on a highly imbalanced dataset shows the ability of the proposed deep multi-modal model to handle this challenge.

\section{Conclusion}
\label{sec:conc}
In this project, a multi-modal cyber-attack detection model is proposed to analyze sensor and network modalities of \ac{ICS} environments and build an abstract joint representation based on these modalities. In the proposed model, a fully connected \ac{DNN} is used to map the sensor modality to the sensor latent space, while a stacked \ac{LSTM} model is used to encode the time-series network data into network latent space. 
The evaluation results show that the proposed model outperformed all other models with 0.99 precision, 0.98 recall, and 0.98 f-measure metrics. Moreover, the proposed model has low false alarms and evaded attack samples that show its ability to handle imbalanced data, which shows the effectiveness of using different modalities in the proposed model.

\bibliographystyle{IEEEtran}
\bibliography{refs}

\begin{thebibliography}{10}
\providecommand{\url}[1]{#1}
\csname url@samestyle\endcsname
\providecommand{\newblock}{\relax}
\providecommand{\bibinfo}[2]{#2}
\providecommand{\BIBentrySTDinterwordspacing}{\spaceskip=0pt\relax}
\providecommand{\BIBentryALTinterwordstretchfactor}{4}
\providecommand{\BIBentryALTinterwordspacing}{\spaceskip=\fontdimen2\font plus
\BIBentryALTinterwordstretchfactor\fontdimen3\font minus
  \fontdimen4\font\relax}
\providecommand{\BIBforeignlanguage}[2]{{%
\expandafter\ifx\csname l@#1\endcsname\relax
\typeout{** WARNING: IEEEtran.bst: No hyphenation pattern has been}%
\typeout{** loaded for the language `#1'. Using the pattern for}%
\typeout{** the default language instead.}%
\else
\language=\csname l@#1\endcsname
\fi
#2}}
\providecommand{\BIBdecl}{\relax}
\BIBdecl

\bibitem{SAKHNINI2021101394}
J.~Sakhnini, H.~Karimipour, A.~Dehghantanha, and R.~M. Parizi, ``Physical layer
  attack identification and localization in cyber–physical grid: An ensemble
  deep learning based approach,'' \emph{Physical Communication}, vol.~47, p.
  101394, 2021.

\bibitem{Zolanvari2019}
M.~Zolanvari, M.~A. Teixeira, L.~Gupta, K.~M. Khan, and R.~Jain, ``{Machine
  Learning-Based Network Vulnerability Analysis of Industrial Internet of
  Things},'' \emph{IEEE Internet of Things Journal}, vol.~6, no.~4, pp.
  6822--6834, 2019.

\bibitem{Zhang2019}
F.~Zhang, H.~A. D.~E. Kodituwakku, J.~W. Hines, and J.~Coble, ``{Multilayer
  Data-Driven Cyber-Attack Detection System for Industrial Control Systems
  Based on Network, System, and Process Data},'' \emph{IEEE Transactions on
  Industrial Informatics}, vol.~15, no.~7, pp. 4362--4369, 2019.

\bibitem{Collier2021}
\BIBentryALTinterwordspacing
K.~Collier, ``{50,000 security disasters waiting to happen: The problem of
  America's water supplies},'' 2021. [Online]. Available:
  \url{https://www.nbcnews.com/tech/security/hacker-tried-poison-calif-water-supply-was-easy-entering-password-rcna1206}
\BIBentrySTDinterwordspacing

\bibitem{9645267}
A.~Al-Abassi, A.~Namavar~Jahromi, H.~Karimipour, A.~Dehghantanha, P.~Siano, and
  H.~Leung, ``A self-tuning cyber-attacks’ location identification approach
  for critical infrastructures,'' \emph{IEEE Transactions on Industrial
  Informatics}, vol.~18, no.~7, pp. 5018--5027, 2022.

\bibitem{Yan2019}
W.~Yan, L.~K. Mestha, and M.~Abbaszadeh, ``{Attack Detection for Securing Cyber
  Physical Systems},'' \emph{IEEE Internet of Things Journal}, vol.~6, no.~5,
  pp. 8471--8481, 2019.

\bibitem{Singh2020}
S.~Singh, H.~Karimipour, H.~HaddadPajouh, and A.~Dehghantanha, \emph{Artificial
  intelligence and security of industrial control systems}.\hskip 1em plus
  0.5em minus 0.4em\relax Cham: Springer International Publishing, 2020, pp.
  121--164.

\bibitem{Paridari2018}
K.~Paridari, N.~O'Mahony, A.~{El-Din Mady}, R.~Chabukswar, M.~Boubekeur, and
  H.~Sandberg, ``{A Framework for Attack-Resilient Industrial Control Systems:
  Attack Detection and Controller Reconfiguration},'' \emph{Proceedings of the
  IEEE}, vol. 106, no.~1, pp. 113--128, 2018.

\bibitem{ShuguangCui2012}
\BIBentryALTinterwordspacing
{Shuguang Cui}, {Zhu Han}, S.~Kar, T.~T. Kim, H.~V. Poor, and A.~Tajer,
  ``{Coordinated Data-Injection Attack and Detection in the Smart Grid: A
  Detailed Look at Enriching Detection Solutions},'' \emph{IEEE Signal
  Processing Magazine}, vol.~29, no.~5, pp. 106--115, sep 2012. [Online].
  Available: \url{http://ieeexplore.ieee.org/document/6279584/}
\BIBentrySTDinterwordspacing

\bibitem{Kurt2018}
\BIBentryALTinterwordspacing
M.~N. Kurt, Y.~Yilmaz, and X.~Wang, ``{Distributed Quickest Detection of
  Cyber-Attacks in Smart Grid},'' \emph{IEEE Transactions on Information
  Forensics and Security}, vol.~13, no.~8, pp. 2015--2030, aug 2018. [Online].
  Available: \url{https://ieeexplore.ieee.org/document/8278264/}
\BIBentrySTDinterwordspacing

\bibitem{7762123}
H.~H. {Pajouh}, R.~{Javidan}, R.~{Khayami}, A.~{Dehghantanha}, and K.~R.
  {Choo}, ``A two-layer dimension reduction and two-tier classification model
  for anomaly-based intrusion detection in iot backbone networks,'' \emph{IEEE
  Transactions on Emerging Topics in Computing}, vol.~7, no.~2, pp. 314--323,
  April 2019.

\bibitem{W19}
S.~{Ponomarev} and T.~{Atkison}, ``Industrial control system network intrusion
  detection by telemetry analysis,'' \emph{IEEE Transactions on Dependable and
  Secure Computing}, vol.~13, no.~2, pp. 252--260, 2016.

\bibitem{karimipourIET}
H.~Karimipour and H.~Leung, ``Relaxation-based anomaly detection in
  cyber-physical systems using ensemble {Kalman} filter,'' \emph{IET
  Cyber-Physical Systems: Theory \& Applications}, 2019.

\bibitem{Krawczyk2016}
B.~Krawczyk, ``{Learning from imbalanced data: open challenges and future
  directions},'' \emph{Progress in Artificial Intelligence}, vol.~5, no.~4, pp.
  221--232, 2016.

\bibitem{sakhnini2019smart}
J.~Sakhnini, H.~Karimipour, and A.~Dehghantanha, ``Smart grid cyber attacks
  detection using supervised learning and heuristic feature selection,'' in
  \emph{{2019 IEEE 7th international conference on smart energy grid
  engineering (SEGE)}}.\hskip 1em plus 0.5em minus 0.4em\relax IEEE, 2019, pp.
  108--112.

\bibitem{DAS2020101935}
T.~K. Das, S.~Adepu, and J.~Zhou, ``Anomaly detection in industrial control
  systems using logical analysis of data,'' \emph{Computers \& Security},
  vol.~96, p. 101935, 2020.

\bibitem{al2020unsupervised}
A.~Al-Abassi, J.~Sakhnini, and H.~Karimipour, ``Unsupervised stacked
  autoencoders for anomaly detection on smart cyber-physical grids,'' in
  \emph{{2020 IEEE International Conference on Systems, Man, and Cybernetics
  (SMC)}}.\hskip 1em plus 0.5em minus 0.4em\relax IEEE, 2020, pp. 3123--3129.

\bibitem{MITTAL202224}
H.~Mittal, A.~K. Tripathi, A.~C. Pandey, M.~D. Alshehri, M.~Saraswat, and
  R.~Pal, ``A new intrusion detection method for cyber–physical system in
  emerging industrial {IoT},'' \emph{Computer Communications}, vol. 190, pp.
  24--35, 2022.

\bibitem{7926429}
Y.~{He}, G.~J. {Mendis}, and J.~{Wei}, ``Real-time detection of false data
  injection attacks in smart grid: A deep learning-based intelligent
  mechanism,'' \emph{IEEE Transactions on Smart Grid}, vol.~8, no.~5, pp.
  2505--2516, 2017.

\bibitem{wang2018}
\BIBentryALTinterwordspacing
H.~Wang, J.~Ruan, G.~Wang, B.~Zhou, Y.~Liu, X.~Fu, and J.~Peng, ``{Deep
  Learning-Based Interval State Estimation of AC Smart Grids Against Sparse
  Cyber Attacks},'' \emph{IEEE Transactions on Industrial Informatics},
  vol.~14, no.~11, pp. 4766--4778, nov 2018. [Online]. Available:
  \url{https://ieeexplore.ieee.org/document/8288611/}
\BIBentrySTDinterwordspacing

\bibitem{Mariam}
M.~M.~N. {Aboelwafa}, K.~G. {Seddik}, M.~H. {Eldefrawy}, Y.~{Gadallah}, and
  M.~{Gidlund}, ``A machine-learning-based technique for false data injection
  attacks detection in industrial iot,'' \emph{IEEE Internet of Things
  Journal}, vol.~7, no.~9, pp. 8462--8471, 2020.

\bibitem{Karimipour2019}
H.~{Karimipour}, A.~{Dehghantanha}, R.~M. {Parizi}, K.~R. {Choo}, and
  H.~{Leung}, ``A deep and scalable unsupervised machine learning system for
  cyber-attack detection in large-scale smart grids,'' \emph{IEEE Access},
  vol.~7, pp. 80\,778--80\,788, 2019.

\bibitem{althobaiti2021intelligent}
M.~M. Althobaiti, K.~P.~M. Kumar, D.~Gupta, S.~Kumar, and R.~F. Mansour, ``An
  intelligent cognitive computing based intrusion detection for industrial
  cyber-physical systems,'' \emph{Measurement}, vol. 186, p. 110145, 2021.

\bibitem{Jahromi2021}
A.~Namavar~Jahromi, H.~Karimipour, A.~Dehghantanha, and K.-K.~R. Choo, ``Toward
  detection and attribution of cyber-attacks in iot-enabled cyber–physical
  systems,'' \emph{IEEE Internet of Things Journal}, vol.~8, no.~17, pp.
  13\,712--13\,722, 2021.

\bibitem{khan2022enhancing}
I.~A. Khan, M.~Keshk, D.~Pi, N.~Khan, Y.~Hussain, and H.~Soliman, ``Enhancing
  {IIoT} networks protection: A robust security model for attack detection in
  internet industrial control systems,'' \emph{Ad Hoc Networks}, vol. 134, p.
  102930, 2022.

\bibitem{majidi2022fdi}
S.~H. Majidi, S.~Hadayeghparast, and H.~Karimipour, ``{FDI} attack detection
  using extra trees algorithm and deep learning algorithm-autoencoder in smart
  grid,'' \emph{{International Journal of Critical Infrastructure Protection}},
  vol.~37, p. 100508, 2022.

\bibitem{jahromi2021deep}
A.~Namavar~Jahromi, H.~Karimipour, and A.~Dehghantanha, ``Deep federated
  learning-based cyber-attack detection in industrial control systems,'' in
  \emph{2021 18th International Conference on Privacy, Security and Trust
  (PST)}.\hskip 1em plus 0.5em minus 0.4em\relax IEEE, 2021, pp. 1--6.

\bibitem{amir2}
A.~NamavarJahromi, H.~Karimipour, and A.~Dehghantanha, ``An ensemble deep
  federated learning cyber-threat hunting model for industrial internet of
  things,'' \emph{Computer Communications}, vol. 198, pp. 108--116, 2023.

\bibitem{KravchikMoshe2021ECAD}
M.~Kravchik and A.~Shabtai, ``\BIBforeignlanguage{eng}{Efficient cyber attack
  detection in industrial control systems using lightweight neural networks and
  pca},'' \emph{\BIBforeignlanguage{eng}{IEEE transactions on dependable and
  secure computing}}, pp. 1--1, 2021.

\bibitem{Goh2017}
J.~Goh, S.~Adepu, K.~N. Junejo, and A.~Mathur, ``A dataset to support research
  in the design of secure water treatment systems,'' in \emph{Critical
  Information Infrastructures Security}, G.~Havarneanu, R.~Setola,
  H.~Nassopoulos, and S.~Wolthusen, Eds.\hskip 1em plus 0.5em minus 0.4em\relax
  Cham: Springer International Publishing, 2017, pp. 88--99.

\end{thebibliography}

\end{document}